\documentclass{moriond}

\def\Journal#1#2#3#4{{#1} {\bf #2}, #3 (#4)}

\def\be{\begin{equation}}
\def\ee{\end{equation}}
\def\bea{\begin{eqnarray}}
\def\eea{\end{eqnarray}}

\usepackage[dvipsnames]{xcolor}
\usepackage{amsmath,amssymb}
\usepackage{physics2}
\usephysicsmodule{ab} 

\makeatletter
\newcommand\vb{\@ifstar\boldsymbol\mathbf}
\newcommand\va[1]{\@ifstar{\vec{#1}}{\vec{\mathrm{#1}}}}
\newcommand\vu[1]{%
\@ifstar{\hat{\boldsymbol{#1}}}{\hat{\mathbf{#1}}}}
\makeatother

\newcommand{\transpose}[1]{\ensuremath{#1^{\mathsf T}}}
\DeclareMathOperator{\diag}{diag}

\newcommand{\Photo}{\includegraphics[height=35mm]{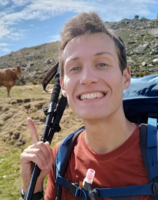}}

\begin{document}
\vspace*{4cm}
\title{An Introduction to Map-making For CMB Experiments}

\author{ Simon Biquard }

\address{Université Paris Cité, CNRS, Astroparticule et Cosmologie\\ F-75013 Paris, France}

\maketitle\abstracts{
    The cosmic microwave background (CMB) anisotropies are a powerful probe of the early universe, and have largely contributed to establishing the current standard cosmological model.
    To extract the information encoded in those tiny variations, one must first compress the raw, time-domain data collected by a telescope into maps of the sky at the observed frequencies, in a procedure known as map-making.
    I provide a general introduction to this problem, and highlight a few specificities of the MAPPRAISER implementation.
}

\section{Formalism}

\subsection{Data model}

The time-ordered data (TOD) collected by the detectors is assumed to have been calibrated and checked for quality during a pre-processing stage.
It can then be used to estimate maps of the sky signal observed at each frequency.
This \emph{map-making} operation represents a significant compression of the data volume, from $\mathcal O(10^{12-14})$ time samples for modern CMB experiments to typically $\mathcal O(10^{5-8})$ sky pixels depending on the particular instrument and sky coverage.
Ideally, this compression should not translate into the loss of any cosmological information, which is why accurate modeling of noise and systematics in the data is required.

The usual data model for the map-making problem links the vector of true sky signal amplitudes $\vb{s}$ and the data vector $\vb{d}$ (a concatenation of the TOD from all detectors) through the application of a \emph{pointing operator} $\vb{P}$ (a tall and skinny matrix which encodes the scanning pattern and orientation of the telescope):
\begin{equation}\label{eq:mm-problem}
    \vb{d} = \vb{P} \vb{s} + \vb{n}
\end{equation}

The time-domain vector $\vb{n}$ represents a stochastic contribution (noise) that vanishes on average.
A given row of the matrix $\vb{P}$ corresponds to the measurement of a pixel $p$ performed by one detector at a given time $t$.
Consequently, each row contains only a small number of non-zero elements, which are determined by the telescope pointing information as well as our knowledge of the instrument.
In most applications, this number is either 1 (for total intensity measurements) or 3 (for polarization-sensitive measurements), which makes the pointing matrix typically very sparse and computationally manageable.
Below are two data model examples for a polarization sensitive bolometer at an angle $\varphi$ with respect to the sky coordinates:
\begin{align}
    d_t &= I_{p_t} + \cos(2\varphi_t)Q_{p_t} + \sin(2\varphi_t)U_{p_t} + n_t
    \\
    d_t &= I_{p_t} + \cos(2\varphi_t + 4\psi_t)Q_{p_t} + \sin(2\varphi_t + 4\psi_t)U_{p_t} + n_t
\end{align}
In the second case, an ideal half-wave plate (HWP) oriented with an angle $\psi$ with respect to the instrument is used to modulate the incoming polarization of the photons.

\subsection{Sky signal estimators}

Formally, map-making is then simply a \emph{linear operation} mapping the measurements $\vb{d}$ to an estimate $\vb{m}$ of the sky amplitudes with some operator $\vb{L}$:
\begin{equation}
    \vb{m} = \vb{L} \vb{d}
\end{equation}

Arguably the simplest possible choice is the \emph{binning} method, $\vb{L} = \pab{\transpose{\vb P} \vb{P}}^{-1} \transpose{\vb P}$, which corresponds to basic averaging of all measurements falling in each pixel.
This is an unbiased estimator which is also easy to compute, but is generally very sub-optimal (noisy), because the correlated time-domain noise is projected onto the sky without any prior weighting.

More generally, we see that requiring the estimator to be unbiased in general constrains the map-making operator to be a left inverse of the pointing matrix:
\begin{equation}
    \aab{\vb m} = \vb{L} \aab{\vb d} = \vb{L} \vb{P} \vb{s} + \underbrace{\vb{L} \aab{\vb n}}_{\text{vanishes}} = \vb{s}
    \quad \Longleftrightarrow \quad
    \vb{L} \vb{P} = \vb{I}
\end{equation}

A family of unbiased estimators is therefore obtained by taking
\begin{equation}\label{eq:mm-operator}
    \vb{L} = \pab{\transpose{\vb P} \vb{W} \vb{P}}^{-1} \transpose{\vb P} \vb{W}
\end{equation}
where the positive definite \emph{weight matrix} $\vb{W}$ determines the resulting noise properties of the maps.
A possible choice is $\vb{W} = \vb{C}^{-1}_{\vb n}$, the inverse of the \emph{noise covariance matrix}.
This corresponds to the generalized least squares (GLS) solution of \eqref{eq:mm-problem} and yields the best (i.e. lowest variance) linear unbiased estimator of $\vb{s}$.
This choice can be interpreted as \emph{maximum likelihood} estimation with the prior that the statistics of the noise vector are Gaussian.

Another possibility is to generalize the weight matrix using a set of temporal modes, or \emph{templates}, which are to be de-projected from the data.
If those time-domain vectors are orthogonal and span the columns of a matrix $\vb{T}$, then the \emph{filtering and weighting} operator $\vb{F}_{\vb T} = \vb{W} \pab{\vb{I} - \vb{T} (\transpose{\vb T} \vb{W} \vb{T})^{-1} \transpose{\vb T} \vb{W}}$ effectively filters all those modes from the data and weights the remaining ones by $\vb{W}$.
That construction allows for unbiased time-domain filtering~\cite{polarbear}.

Note, for completeness, that one may also compute a biased estimate of the sky, for example of the form $\vb{\tilde{m}} = \pab{\transpose{\vb P} \vb{\Lambda} \vb{P}}^{-1} \transpose{\vb P} \vb{F} \vb{d}$, where $\vb{F}$ is a filtering operator and $\vb{\Lambda}$ a different weight matrix (typically diagonal), which makes the computation much cheaper.
This approach is called \emph{filter-and-bin} map-making and has been successfully used in many experiments, such as BICEP/Keck~\cite{bicep2-ii}.
The bias must be corrected for at a later stage of the analysis, typically by using an estimated transfer function between the estimate and the true sky.

\subsection{Computational aspects}
\label{sec:computational}

Because of the size of the data involved, the map-making problem is typically tackled using \emph{iterative methods}.
Indeed, direct methods that compute some decomposition of the system matrix generally scale as the cube of the system size, which makes them prohibitively expensive for large datasets.
In other words, the $\pab{\transpose{\vb P} \vb{W} \vb{P}}^{-1}$ kernel of \eqref{eq:mm-operator}, which can be a dense matrix, is never explicitly computed, and instead the estimate is obtained by solving for $\vb{m}$ the map-making equation:
\begin{equation}
    \pab{\transpose{\vb P} \vb{W} \vb{P}} \vb{m} = \transpose{\vb P} \vb{W} \vb{d}
\end{equation}
in a way that only requires matrix-vector operations.
The multiplication of the system matrix $\vb{A} = \transpose{\vb P} \vb{W} \vb{P}$ with a vector $\vb{x}$ is done from right to left in a series of three products that can be performed efficiently.

Because the system matrix is symmetric and positive definite, conjugate gradient techniques are appropriate for this task and have been widely used in the map-making context.
To speed up the convergence of the solver, a \emph{preconditioner} is often employed: its role is to reduce the condition number of the system.
This approach is called the Preconditioned Conjugate Gradient (PCG) method.
A preconditioner $\vb{M}$ should approximate the inverse of $\vb{A}$, such that solving the preconditioned system with matrix $\vb{M} \vb{A}$ is faster than  the original one.
For the GLS solution with $\vb{W} = \vb{C}^{-1}_{\vb n}$, a possible choice of preconditioner which can be easily computed and stored in memory is the block-diagonal matrix $\vb{M}_{bd} = \pab{\transpose{\vb P} \diag(\vb{C}^{-1}_{\vb n}) \vb{P}}^{-1}$.

\section{The MAPPRAISER library}

In this section, I summarize the main features of the MAPPRAISER library, but point the reader to the release paper~\cite{mappraiser} and references therein for a more detailed and thorough description.

\subsection{Overview}

This software is built around two libraries written in C, available \href{https://github.com/B3Dcmb/midapack}{on GitHub}.
The first one provides low-level operations such as sparse linear algebra for pointing and template operators and custom communication schemes for distributed data reduction.
The second one builds on the first one and implements different numerical techniques and map-making methods to compute unbiased estimates of the sky signal.
Both libraries use the MPI programming model in order to operate on distributed computing systems and make use of a large number of compute nodes.
Finally, a Python wrapper allows the map-making code to be used with the TOAST software package, a simulation and analysis tool for time-ordered data.

The MAPPRAISER library implements two specific, optimized versions of the operator $\vb{L}$ introduced in \eqref{eq:mm-operator}.
\begin{enumerate}
    \item the GLS solution $\vb{W} = \vb{C}^{-1}_{\vb n}$, assuming that the noise is close to piece-wise stationary and uncorrelated between detectors: the noise covariance matrix has a block-diagonal Toeplitz structure, which allows for efficient matrix-vector operations using fast Fourier transforms.
    \item the \emph{templates} method, using the aforementioned filtering and weighting operator $\vb{F}_{\vb T} = \vb{W} \pab{\vb{I} - \vb{T} (\transpose{\vb T} \vb{W} \vb{T})^{-1} \transpose{\vb T} \vb{W}}$, allowing for arbitrary time-domain templates but diagonal (uncorrelated) noise weights.
\end{enumerate}

\subsection{Preconditioners}

One of the flagship features of the MAPPRAISER library is the implementation of two preconditioning options for the PCG solver. The first option is the standard Block-Jacobi preconditioner $\vb{M}_{bd}$ introduced in \autoref{sec:computational}, computed using the diagonal of the inverse noise covariance matrix.
It is block-diagonal, with one square block per sky pixel, whose size is equal to the number of non-zero elements in a row of the pointing matrix.

\begin{figure}
\centering
\includegraphics[width=0.55\textwidth]{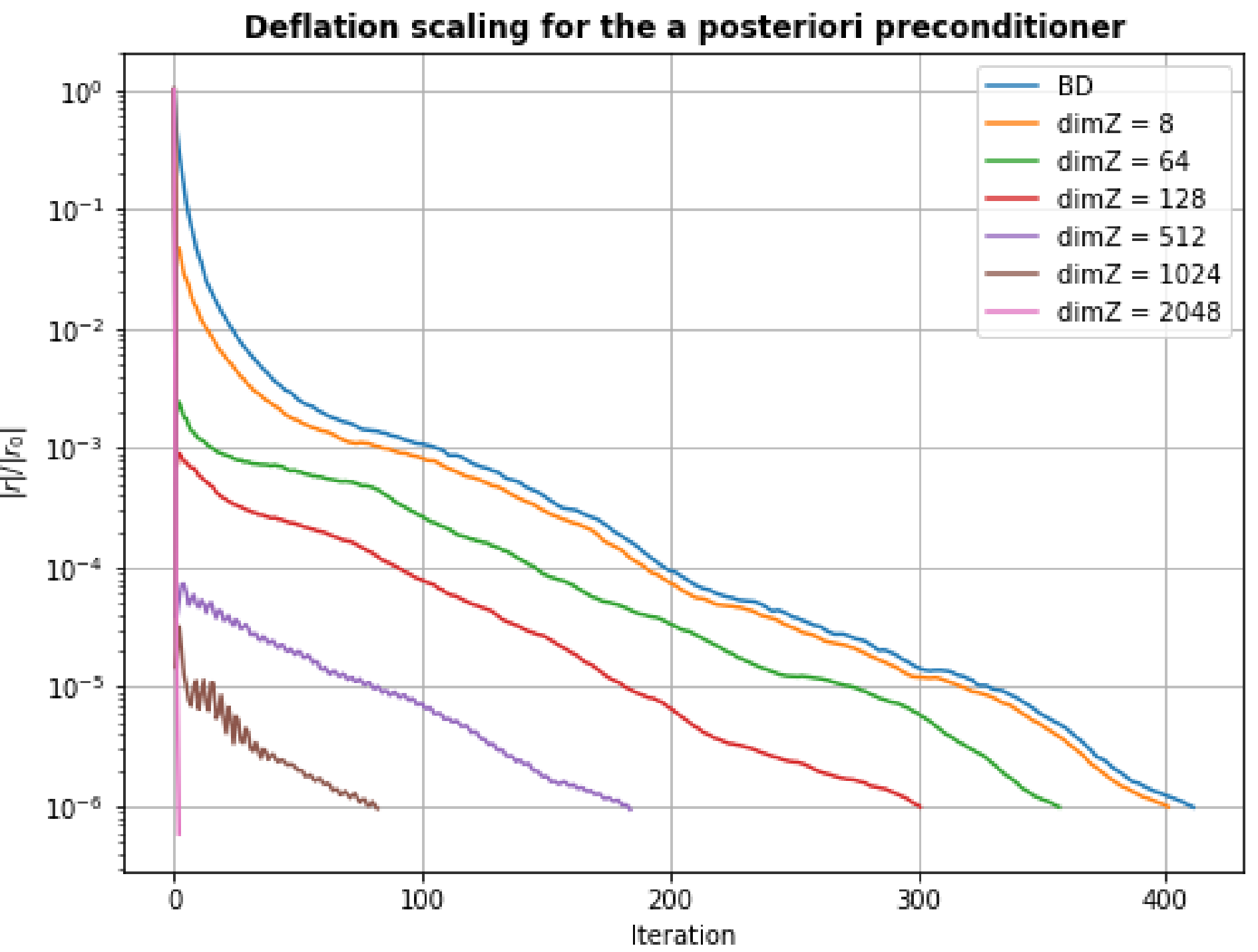}
\caption{Relative reduction of residuals as a function of the iteration number for the PCG solver (convergence is defined by reaching a $10^{-6}$ reduction factor).
The Block-Jacobi preconditioner $\mathbf{M}_{bd}$ (in blue) is compared to the a posteriori two-level preconditioner $\mathbf{M}_{2lvl}$ (other colors) for different dimensions of the deﬂation subspace.
When $\dim{\mathbf Z} = 2048$, convergence is reached in one iteration.
Figure taken from the MAPPRAISER release paper.
}
\label{fig:2lvl_iter}
\end{figure}

The second option is the \emph{two-level preconditioner}.
The first level uses the deflation technique and aims at removing a subspace containing the smallest eigenvalues of the system matrix.
The second level adds a correction to the deflated matrix which shifts the eigenvalues of the subspace towards unity, effectively reducing the condition number of the system matrix.
The preconditioner is formally defined as:
\begin{align}
    \vb{M}_{2lvl} &= \overbrace{\vb{M}_{bd} \pab{\vb{I} - \vb{A} \vb{Q}}}^{\text{1st level}} + \overbrace{\vb{Q}}^{\text{2nd level}}
    \\
    \vb{Q} &= \vb{Z} \pab{\transpose{\vb Z} \vb{A} \vb{Z}}^{-1} \transpose{\vb Z}
\end{align}
where $\vb{Z}$ is the deflation subspace matrix (its pixel-domain columns span a subspace to be suppressed).
The key element here is the construction of $\vb{Z}$.
In our tests, the most effective way of doing that has proven to be the so-called \emph{a posteriori} construction, where $\vb{Z}$ is built by estimating the relevant eigenpairs of the system matrix $\vb{A}$ with the Lanczos algorithm.
The performance of this preconditioner is illustrated in Figure~\ref{fig:2lvl_iter}.

\section{Conclusion and future perspectives}

The two-level preconditioning approach effectively solves the map-making problem from a purely \emph{computational} point of view.
Indeed, the precomputation cost associated to a large deflation subspace can be compensated by reusing it for solving many similar systems in a row (e.g. in Monte Carlo simulations).
While I expect this technique to be useful in the context of increasing data volumes, it remains to be seen whether other aspects of the problem such as I/O will not become limiting factors.

SB acknowledges funding from the European Research Council (ERC; SCIPOL\footnote{\url{https://scipol.in2p3.fr}} project) under the European Union’s Horizon 2020 research and innovation program (Grant agreement No. 101044073).

\end{document}